\documentclass[12pt]{article}
\usepackage[utf8]{inputenc}
\usepackage{textcomp}
\usepackage{mathtools}
\usepackage{bm}
\usepackage[dvipsnames]{xcolor}
\usepackage{graphicx}
\usepackage{dcolumn}
\usepackage{jheppub}
\usepackage{iip-arxiv}

\makeatletter
\pdfoutput=1

\DeclareMathAlphabet{\mathbbold}{U}{bbold}{m}{n}

\newcommand{\poisson}[3][]{\left\{ #2, #3\right\} }

\newcommand{\afltbasis}[1][\vec{\lambda}]{|P\rangle_{#1}}

\preprint{CECS-PHY-18/02, ITEP-TH-20/18 }

\title{Lifshitz Scaling, Microstate Counting from Number Theory and Black
Hole Entropy}

\author[a,b]{Dmitry Melnikov,}  
\author[a]{Fábio Novaes,}
\author[c]{Alfredo P\'erez}
\author[c]{and Ricardo Troncoso}

\affiliation[a]{International Institute of Physics, Federal University of Rio Grande do Norte,\\
Av. Odilon Gomes de Lima 1722, Capim Macio, Natal-RN 59078-400,
Brazil}
\affiliation[b]{Institute for Theoretical and Experimental Physics,\\ B. Cheremushkinskaya 25, Moscow 117218, Russia}
\affiliation[c]{Centro de Estudios Científicos (CECs), Av. Arturo Prat 514,  Valdivia, Chile}

\email{dmitry at iip.ufrn.br}
\email{fnovaes at iip.ufrn.br}
\email{aperez at cecs.cl}
\email{troncoso at cecs.cl}

\abstract{Non-relativistic field theories with anisotropic scale
  invariance in (1+1)--d are typically characterized by a dispersion
  relation $E\sim k^{z}$ and dynamical exponent $z>1$.  The asymptotic
  growth of the number of states of these theories can be described by
  an extension of Cardy formula that depends on $z$. We show that this
  result can be recovered by counting the partitions of an integer
  into $z$--th powers, as proposed by Hardy and Ramanujan a century
  ago.  This gives a novel duality relationship between the characteristic
  energy of the dispersion relation with the cylinder radius and the
  ground state energy. For free bosons with Lifshitz scaling,
this relationship is shown to be identically fulfilled by virtue of the reflection
property of the Riemann $\zeta$-function. The quantum
Benjamin-Ono$_{2}$ (BO$_{2}$) integrable system, relevant in the AGT
correspondence, is also analyzed. As a holographic realization, we provide
a special set of  boundary conditions for which the reduced
  phase space of Einstein gravity with a couple of $U\left(1\right)$
  fields on AdS$_{3}$ is described by the BO$_{2}$ equations. This
  suggests that the phase space can be quantized in terms of quantum
  BO$_{2}$ states. Indeed, in the semiclassical limit, the ground state energy
  of BO$_{2}$ coincides with the energy of global AdS$_{3}$, and the
  Bekenstein-Hawking entropy for BTZ black holes is recovered from the
  anisotropic extension of Cardy formula.}

\makeatother

\begin{document}
\maketitle

\section{Introduction}

\label{sec:introduction}

Non-relativistic field theories in $(1+1)$--dimensions, possessing
anisotropic Lifshitz scaling of the form 
\begin{equation}
t\rightarrow\lambda^{z}t\,,\,x\rightarrow\lambda x,\label{eq:Lifshitz-scaling}
\end{equation}
are typically characterized by modes with a dispersion relation $E\sim k^{z}$
and entropy $S\sim E^{\frac{1}{z+1}}$, with dynamical exponent $z>1$.
They have been extensively studied in the context of the holographic
AdS/CMT correspondence, see e.g. \cite{Taylor:2008tg,Hartnoll:2009sz,Hartnoll:2011fn,Taylor:2015glc}.
A number of condensed matter systems are known to enjoy this type
of scaling. For instance, the quantum Hall fluid has been proposed
to possess a non-linear dispersion with $z=2$ \cite{Bettelheim2006,Wiegmann:2011aa},
while the (1+1)-dimensional Bose gas in cold atom systems can be related
to $z=3$ dispersion~\cite{Sotiriadis2017}. The non-relativistic effective
field theory description of such systems is invariant under the \emph{Lifshitz
group}, which in (1+1)-d is generated by translations in space and
time and the anisotropic scale transformation \eqref{eq:Lifshitz-scaling}.
At finite temperature, chiral movers with Lifshitz scaling can be
described in terms of the torus partition function 
\begin{equation}
Z[\tau;z]=\sum_{E}\rho_{z}(E)e^{2\pi\ell i\tau E},\label{eq:lifshitz-generic-partition}
\end{equation}
where $\tau=i\beta/\left(2\pi\ell\right)$ is the modular parameter,
defined in terms of the inverse temperature $\beta$ and the radius
of the cylinder $\ell$, and $\rho_{z}(E)$ is the density of states
at fixed energy $E$. As explained in \cite{Gonzalez2011,Perez2016},
assuming modular invariance of a well-defined partition function
$Z[\tau; z]$ in the anisotropic
case implies that 
\begin{equation}
Z[\tau;z]=Z[i^{1+\frac{1}{z}}\tau^{-\frac{1}{z}};z^{-1}].\label{eq:anisotropic-modular-inv}
\end{equation}
 Thus, eq.~\eqref{eq:anisotropic-modular-inv} can be regarded as the
  defining property of the partition functions of the  theories we are interested in.
  Note that in the case of isotropic scaling ($z=1$), eq. 
  \eqref{eq:anisotropic-modular-inv} reduces
  to the well--known modular invariance in CFT$_{2}$ \cite{DiFrancesco1997a}.

  In the partition function $Z[\tau;z]$, the modular parameter of the torus
  $\tau$ plays the standard role as a chemical 
  potential, while the dynamical exponent $z$ turns out to be a parameter without 
  variation which possesses a well-defined transformation property under
  a modular transformation\footnote{Note that $z$ plays a similar role as that of
      the parameter $\lambda$ that characterizes the $T\bar{T}$ deformations of CFTs 
      (see e.g. \cite{Zamolodchikov:2004ce, Smirnov:2016lqw, Cavaglia:2016oda, McGough:2016lol,Kraus:2018xrn}). Indeed,
      $\lambda$ is not varied in the partition function and it also possesses a precise 
      transformation property under modular transformations \cite{Datta:2018thy,Aharony:2018bad}. Modular
      invariance of the original CFT$_2$ is recovered for $\lambda=0$, while in our 
      case it does for $z=1$.}.

It is worth highlighting that in eq.~\eqref{eq:anisotropic-modular-inv}, the existence 
of two different 
theories with different dynamical exponents ($z$ and $1/z$) that map into each 
other under a modular transformation is implicitly assumed. Thus, modular invariance 
of the partition function in the anisotropic case, as in eq.~\eqref{eq:anisotropic-modular-inv},
relates the high and low energy spectrum of the corresponding Hamiltonians.

If one further assumes that the spectrum of the theory described by a dynamical  
exponent $z$ possesses a gap with a non-vanishing ground state energy
given by $-E_{0}[z]$, then the asymptotic growth of the number of
states at fixed energy $E\gg\left|E_{0}[z]\right|$ can be obtained
from the inverse Laplace transform of \eqref{eq:anisotropic-modular-inv} in
the steepest descent approximation 
\begin{equation}
\rho_{z}(E)\approx\,\exp\left[2\pi\ell(1+z)\left(\frac{\left|E_{0}[z^{-1}]\right|}{z}\right)^{\frac{z}{1+z}}E^{\frac{1}{1+z}}\right],\label{eq:anisotropic-growth-states}
\end{equation}
and hence 
\begin{equation}
S=\log\rho_{z}(E)=2\pi\ell(1+z)\left(\frac{\left|E_{0}[z^{-1}]\right|}{z}\right)^{\frac{z}{1+z}}E^{\frac{1}{1+z}},\label{eq:leading-entropy}
\end{equation}
stands for the leading term of the microcanonical entropy.
Note that
for $z=1$, the entropy reduces to the well-known Cardy formula in
CFT$_{2}$ \cite{Cardy1986a}. The logarithmic correction to \eqref{eq:leading-entropy}
was discussed in \cite{Grumiller2017} (see also \cite{Shaghoulian2015}).

It is worth emphasizing that the high/low temperature duality 
of the partition function expressed by
 \eqref{eq:anisotropic-modular-inv} can be argued to emerge from 
 the combination of two purely geometric properties that a generic field theory with 
anisotropic scaling defined on a torus should possess. 
Indeed, on one hand, the lattice that corresponds to a generic torus  
is invariant under S-duality which swaps both periods preserving the orientation (see e.g. \cite{apostolmodular}).
Besides, Lifshitz algebras in 2d with dynamical exponents $z$ and $1/z$ turn out to be 
isomorphic, since they are related by a change of basis in which the generators of space 
and Euclidean time translations are swapped \cite{Gonzalez2011}. Hence, the partition function of
 a theory possessing Lifshitz scaling that can be
 consistently defined on a torus should be, in particular, invariant under the combined
 action of S-duality and the isomorphism aforementioned. Consistency of both operations
 then provides evidence about
 the existence of a suitable dual theory
 described by a dynamical exponent $z^{-1}$, so that the partition
 function could be assumed to possess the duality property in 
 eq. \eqref{eq:anisotropic-modular-inv}. In turn, it would be interesting to 
 explore the possibility\footnote{We thank an anonymous referee 
for this suggestion.} that the high/low temperature 
duality relationship applied to the theories described by $z>1$ could actually be used 
 to define the corresponding dual theories with 
dynamical exponent $z^{-1}$.
  
 Hitherto, strong support to this high/low temperature 
 duality can be gathered from a number of holographic examples, formulated in terms of 
 different gravitational theories in three spacetime dimensions. In all cases, the 
 black hole entropy, which is not necessarily given by a quarter of the event horizon 
 area, is precisely recovered from \eqref{eq:leading-entropy} 
 provided that $E$ and $E_0$ correspond to the (left/right) energies of the black 
 hole and the ground state, respectively. The ground state configuration
 is given by a soliton, which turn out to be diffeomorphic to the black hole provided that
 the modular parameters of the corresponding tori at the boundary are related through
 the combination of S-duality and anisotropic scaling, precisely as in  
 \eqref{eq:anisotropic-modular-inv}. Different cases of asymptotically Lifshitz 
 black holes and their corresponding solitons were discussed in \cite{Gonzalez2011}
 (see also \cite{AyonBeato:2009nh}), and in
  \cite{Ayon-Beato:2015jga, Bravo-Gaete:2015iwa}; while a different class of examples
  in which the anisotropic scaling is induced by special choices of boundary conditions 
  were discussed in \cite{Perez2016} and in \cite{Afshar2017,Grumiller2017}
   (see also \cite{Afshar2016}).

   The appearance of the absolute value of the ground state energy
   $|E_{0}[z^{-1}]|$ in \eqref{eq:anisotropic-growth-states} and
   \eqref{eq:leading-entropy}, which precisely means taking the
   absolute value of the ground state energy $-E_{0}[z]$ and then
   evaluating it at $z^{-1}$, deserves some explanation. As discussed
   in \cite{Perez2016}, for certain special values of $z$, e.g.
   $z=4n-1$, the (Lorentzian) ground state energy turns out to be
   positive.  In these cases, in the Euclidean continuation, the
   corresponding holographic black hole configuration becomes
   diffeomorphic to the solitonic ground state, but with reversed
   orientation. This is due to the fact that the lapse function of an
   ADM foliation of the spacetime that describes the ground state
   reverses its sign. From the point of view of a thermal field theory
   in 2D with anisotropic Lifshitz scaling, that is defined on a
   torus, the lapse function can always be reabsorbed by the modular
   parameter (see e.g. \cite{Henneaux:2013dra,Bunster:2014mua}).
   Thus, in these special cases, the high temperature configuration is
   also expected to be related to the ground state through
   S-duality. The direct implementation of this latter relationship in
   the 2D field theory is not entirely clear, but nonetheless, being
   inspired from the results in the bulk, the procedure can be
   interpreted in the following way: in the special cases, the
   S-modular transformation should incorporate some additional
   suitable operation that reverses the sign of the thermal period,
   which then also has the effect of flipping the sign of the ground
   state energy. Thus, in these cases, the suitable quantity that
   enters into \eqref{eq:anisotropic-growth-states} and
   \eqref{eq:leading-entropy} is $\left|E_{0}[z^{-1}]\right|$.  This
   subtlety can be particularly well visualized in a concrete example
   we provide in sec.~\ref{sec:Free boson} when $z=2m+1$.  Indeed, for
   odd values of $m$, the ground state energy determined by
   $-E_{0}[z]$ in \eqref{eq:groundstateboson}, turns out to be
   manifestly positive.

In the next section, we provide further evidence for the validity of this 
high/low temperature duality from a completely different approach, in which the 
microscopic counting of the states is performed applying some not so well-known
 results from number theory, developed a century ago by Hardy and Ramanujan. 
 Consistency then provides a novel and 
nontrivial duality relation between the characteristic energy of the dispersion relation
 with the cylinder radius and the ground state energy. 
 
 In section \ref{sec:Free boson}, we consider free bosons with Lifshitz scaling 
 in 2d, and we show that the duality relationship aforementioned 
 turns out to be identically fulfilled, in a nontrivial way, by virtue of the reflection
property of the Riemann $\zeta$-function. Thus, this example provides strong evidence
supporting the fact that the high/low temperature duality can be explicitly realized directly in 2d 
without the need of holographic setups.

Section \ref{sec:micr-count-quant} is devoted to the analysis of the quantum Benjamin-Ono$_2$
 (BO$_2$)
integrable system, which possesses anisotropic Lifshitz scaling and also turns out to be
 relevant in the AGT correspondence. In particular, we show that in the semiclassical limit, for
$z$ even, the spectrum becomes dominated by descendants, so that the characteristic
energy can be precisely identified, and hence, the entropy can be obtained along 
the lines of number theory. In the case of $z$ odd, one is able to obtain the value of the 
ground state energy, and the entropy is consequently obtained by means of the 
anisotropic extension of Cardy formula. 

Finally, relying on the results mentioned above, in section \ref{sec:geom-benj-ono_2}
we provide an explicit holographic realization of BO$_2$, in which the 
semiclassical limit can be suitably taken. Concretely, we propose a new 
set of boundary conditions for which the reduced phase space of Einstein gravity with a 
couple of $U(1)$ fields on AdS$_3$ turns out to be described by the BO$_2$ equations. 
Noteworthy, this example suggests that the phase space of the gravitational theory
in the bulk can be quantized in terms of quantum BO$_2$ states. Indeed, the ground state energy
 of quantum BO$_2$, in the semiclassical limit, is shown to exactly coincide with the energy
 of global AdS$_3$. This fact allows to precisely recovering the Bekenstein-Hawking entropy
  for BTZ black holes in terms of the anisotropic extension of Cardy formula in
   \eqref{eq:leading-entropy}. 

\section{Microstate Counting from Number Theory}

\label{sec:Microstate-counting-from}

In this section, we show that for (1+1)--dimensional weakly-coupled
systems at high temperatures on a cylinder of radius $\ell$, the
leading term of the asymptotic growth of the number of states $\rho_{z}\left(E\right)$
in \eqref{eq:leading-entropy} agrees with the asymptotic growth of
the number of partitions of an integer $N$ into $z$-th powers $p_{z}(N)$.
This is so provided that the ground state energy and the radius
of the cylinder are precisely linked with the characteristic energy
of the quasiparticles. Indeed, if the interactions are weak enough
so that at high temperature the system behaves as a gas of free quasiparticles,
as pointed out in the introduction, the dispersion relation has to
be of the form 
\begin{equation}
E_{n}=\varepsilon_{z}\left(k_{n}\ell\right)^{z}=\varepsilon_{z}n^{z},\label{eq:disp}
\end{equation}
where $k_{n}=n/\ell$ is the momentum, $n$ is a non-negative
integer and $\varepsilon_{z}$ stands for the characteristic energy
of the modes. The total energy is then given by 
\[
E=\sum_{i}E_{n_{i}}=\varepsilon_{z}\sum_{i}n_{i}^{z}=\varepsilon_{z}N.
\]
Therefore, assuming the ordering $n_{1}\geq n_{2}\geq\ldots\geq0$
to count only indistinguishable configurations, the number of states
with fixed energy $E$ corresponds to the combinatorial problem of
finding the number of \emph{power partitions} $p_{z}(N)$ for fixed
$N=\sum_{i}n_{i}^{z}=E/\varepsilon_{z}$. Quite remarkably, this problem
was solved in 1918 by Hardy and Ramanujan \cite{hardy1918}. Indeed,
in one of the last formulas of their paper, one finds that for large
$N$, the leading term of the asymptotic growth of power partitions
is given by 
\begin{equation}
p_{z}(N)\approx\exp\left[(1+z)\left(\frac{\Gamma\left(1+\frac{1}{z}\right)\zeta\left(1+\frac{1}{z}\right)}{z}\right)^{\frac{z}{1+z}}N^{\frac{1}{1+z}}\right].\label{eq:asymptotic-power-partitions}
\end{equation}
Surprisingly, for a generic $z>1$ the result was actually a conjecture, proven
later by Wright in 1934 using generalized Bessel functions \cite{wright1934asymptotic}.
A simplified proof has been recently given in \cite{Vaughan2015}
for $z=2$, and extended to $z\geq2$ in \cite{Gafni2016}, both using
the Hardy-Littlewood circle method. Hence, at high temperature, the
leading term of the entropy can be read from \eqref{eq:asymptotic-power-partitions},
and it is given by 
\begin{equation}
S=\log p_{z}\left(N\right)=(1+z)\left(\frac{\Gamma\left(1+\frac{1}{z}\right)\zeta\left(1+\frac{1}{z}\right)}{z}\right)^{\frac{z}{1+z}}\left(\frac{E}{\varepsilon_{z}}\right)^{\frac{1}{1+z}}.\label{eq:Sepsilon}
\end{equation}
One then concludes that at high temperature, the asymptotic growth
of the number of states obtained from anisotropic modular invariance,
given by $\rho_{z}\left(E\right)$ in \eqref{eq:anisotropic-growth-states},
agrees with the one from number theory given by $p_{z}\left(N\right)$
in \eqref{eq:asymptotic-power-partitions}, provided that the characteristic
energy of the dispersion relation is related to the radius of the
cylinder and the non-vanishing ground state energy according to 
\begin{equation}
\left(\varepsilon_{\frac{1}{z}}\right)^{z}=\frac{\Gamma(1+z)\zeta(1+z)}{\left(2\pi\ell\right)^{1+z}}\frac{1}{|E_{0}[z]|}.\label{eq:epsilon0}
\end{equation}

Note that,
for $z=1$, the characteristic energy is related to the \emph{effective}
central charge as $c_{\text{eff}}=\left(\varepsilon_{1}\ell\right)^{-1}$,
so that according to \eqref{eq:epsilon0}, the energy of the ground
state acquires the expected form for chiral movers 
\begin{equation}
\left|E_{0}\left[1\right]\right|=\frac{c_{\text{eff}}}{24\ell}=\frac{1}{24\ell^{2}\varepsilon_{1}}.\label{eq:E0CFT}
\end{equation}

It is worth emphasizing that the ability to express the leading term of the entropy
 in terms of the characteristic
energy as in \eqref{eq:Sepsilon} possesses an advantage, since it opens up the possibilities to
perform the microscopic counting even if the ground state energy vanishes.

 Besides, if the ground state energy does not vanish, expressing the entropy as
  in \eqref{eq:leading-entropy} certainly helps, since its value can be directly obtained
  in cases where the microscopic counting cannot
be explicitly performed.

Therefore, by virtue of \eqref{eq:epsilon0}, the high/low temperature duality
 for systems with anisotropic scaling not only acquires additional support, but it
 becomes enhanced from solid results in number theory.

An additional interesting remark is in order. Note that the asymptotic growth of the number
of states in \eqref{eq:anisotropic-growth-states}, which was obtained from
modular invariance in the anisotropic case, actually holds
for arbitrary real values of $z>0$ \footnote{As pointed out in
 \cite{Perez2016,Afshar2017,Grumiller2017}, 
in the
limit $z\rightarrow0$ there is also a very intriguing link with the results
in \cite{Afshar2016} about ``soft hair'' in the sense of Hawking,
Perry and Strominger \cite{Hawking2016,Hawking2017}.}. Therefore, by virtue of the equivalence
of both ways of computing the entropy, expressed in \eqref{eq:leading-entropy}
and \eqref{eq:Sepsilon}, respectively, one is naturally led to conjecture that
the expression for the asymptotic growth of the power partitions of
Hardy and Ramanujan can actually be extended to hold for positive
real values of $z$. The partition problem in this case can be then naturally defined as 
$E/\varepsilon_{z}=N=\sum_{i}\left\lfloor n_{i}^{z}\right\rfloor$,
where $\left\lfloor x \right\rfloor$ 
stands for the floor of $x$. Indeed, very recent results in number theory
give support to this conjecture, since it has already been proved
for $z=\frac{1}{2}$ in \cite{luca2016explicit} and for $0<z<1$
in \cite{li2018r}. Remarkably, Li and Chen in \cite{li2018r} have
also arrived to the same conjecture, but following a completely different
line of reasoning.

\section{Free Boson with Lifshitz Scaling}

\label{sec:Free boson}

In order to test the results of the previous section, it is instructive
to consider the simple case of a free boson with Lifshitz scaling  \cite{lifshitz1941theory} (see also e.g. \cite{Hartnoll:2009sz,Taylor:2015glc}),
described by

\begin{equation}
I=\frac{1}{2}\int dtdx\left[\left(\partial_{t}\varphi\right)^{2}-\sigma^{2\left(z-1\right)}\left(\partial_{x}^{z}\varphi\right)^{2}\right],\label{eq:ActionFreeBoson}
\end{equation}
with $0\leq x<2\pi\ell$. Here $\sigma$ is an arbitrary parameter
with unit of length, and the units have been chosen such that, for
$z=1$, the speed of light is unity. The dispersion relation of the
modes then reads 
\begin{equation}
E_{n}=\pm\varepsilon_{z}\left|n\right|^{z},\quad\varepsilon_{z}=\frac{\sigma^{z-1}}{\ell^{z}},\label{eq:free-boson-dispersion}
\end{equation}
so that the characteristic energy of left and right movers matches
eq. \eqref{eq:disp}. The Hamiltonian of chiral movers can then be
written as 
\begin{align}
H_{z}=2\frac{\sigma^{z-1}}{\ell^{z}}\sum_{n>0}n^{z-1}a_{-n}a_{n}-E_{0}[z],\label{eq:free-bosons}
\end{align}
where $[a_{n},a_{m}]=\frac{n}{2}\delta_{n,-m}$, and by virtue of
$\zeta$-function regularization, the ground state energy is determined by
\begin{equation}
E_{0}[z]=-\frac{1}{2}\frac{\sigma^{z-1}}{\ell^{z}}\zeta(-z).\label{eq:groundstateboson}
\end{equation}
Note that when $z$ takes odd values, the ground state energy \eqref{eq:groundstateboson}
becomes non-trivial, and remarkably, the duality relation between
the characteristic energy $\varepsilon_{z}$, the energy of the ground
state $E_{0}\left[z\right]$ and the radius of the cylinder $\ell$
in \eqref{eq:epsilon0} becomes identically fulfilled by the reflection
property of the Riemann $\zeta$-function 
\[
\zeta\left(-z\right)=-\frac{1}{2^{z}\pi^{1+z}}\sin\left(\frac{\pi z}{2}\right)\Gamma\left(1+z\right)\zeta\left(1+z\right).
\]

Hence, for odd values of $z$, the leading term of the entropy
can be either obtained from the number theory counting as in  \eqref{eq:Sepsilon} 
with $\varepsilon_{z}=\frac{\sigma^{z-1}}{\ell^{z}}$,
or equivalently by virtue of the expression obtained from the high/low temperature duality in 
\eqref{eq:leading-entropy} with $E_{0}\left[z\right]$
given by \eqref{eq:groundstateboson}.

Therefore, the free boson with Lifshitz scaling for odd values of $z$ certainly appears to be
an explicit example for which the high/low temperature duality can be manifestly realized 
in 2d.

For the case of even values of $z$, an interesting remark is in order. Indeed, in 
this case, one of the hypotheses assumed in order to derive the asymptotic 
growth of the number of states from modular invariance in the anisotropic case 
in \eqref{eq:leading-entropy} is not fulfilled, since the ground state energy in
 \eqref{eq:groundstateboson} manifestly vanishes for $z=2n$.
  Nonetheless, in this case one is able to precisely identify the characteristic energy 
  of the dispersion relation $\varepsilon_{z}$ 
 to be given by \eqref{eq:free-boson-dispersion}, and hence, the entropy can
  still be directly obtained from the number theory counting given by  
  \eqref{eq:Sepsilon}.
  
For a generic value of $z$, it is also worth pointing out that according to number theory,
 the sequence
 of power partitions
possesses the following generating function (see e.g. \cite{hardy1918,Gafni2016})
\[
\sum_{N=0}^{\infty}p_{z}(N)q^{N}=\prod_{n=1}^{\infty}\frac{1}{1-q^{n^{z}}},
\]
and hence, the partition function for free bosons with Lifshitz scaling
acquires the form 
\begin{equation}
Z[\tau;z]={\cal N}\left(\tau;z\right)\left|q^{-E_{0}[z]}\prod_{n=1}^{\infty}\frac{1}{1-q^{n^{z}}}\right|^{2},\label{eq:1-1-1}
\end{equation}
with $q=e^{2\pi i\tau}$. Here ${\cal N}\left(\tau;z\right)$ stands
for a non-exponential factor coming from the contribution
of zero modes, so that it does not modify the leading high temperature
asymptotics of $Z[\tau;z]$. Subleading corrections and further details
about its precise form in connection with modular invariance in the anisotropic 
case will be addressed in \cite{WIP}.

Note that for $z=1$, the action \eqref{eq:ActionFreeBoson} reduces
to the one of a free boson in CFT\emph{$_{2}$}, while the energy
of the ground state energy is recovered from \eqref{eq:groundstateboson}
to be given by $-E_{0}[1]=-\frac{1}{24\ell}$, in agreement with the
known result for chiral movers. Thus, the duality relation in \eqref{eq:epsilon0}
reduces to \eqref{eq:E0CFT}, which is consistent with the fact that
$c_{\text{eff}}=1$. The suitable factor of the partition function in this
case is given by ${\cal N}\left(\tau;1\right)=\text{Im}\left(\tau\right)^{-1/2}$
(see e.g. \cite{DiFrancesco1997a}).

As an ending remark of this section, we would like to emphasize that the analysis 
of the free boson with Lifshitz scaling performed here, including its 
statistical mechanics, as far as the authors' knowledge, is new in the literature.
\section{Microstate Counting and the Quantum Benjamin-Ono$_{2}$ Hierarchy}

\label{sec:micr-count-quant}

In the previous section, we discussed a simple free bosonic model
with Lifshitz scaling $z$ and how its spectrum is connected to the
partitions of integers into $z$-th powers. Here we describe a quantum
integrable system, with an infinite set of conserved quantities, presenting
Lifshitz scaling in the semiclassical limit, the quantum Benjamin-Ono$_{2}$ model. 
This will give an interesting link between the
semiclassical limit of quantum systems, microstate counting of models
with Lifshitz scaling and gravitation on AdS$_{3}$.

\subsection{Classical Formulation of the BO$_{2}$ Hierarchy}

\label{sec:classical-bo_2}

The Benjamin-Ono equation describes deep inner waves in a stratified
fluid, being then a counterpart of the KdV equation for propagation
in a shallow depth channel \cite{matsuno1984bilinear}. Both equations
possess solitonic solutions and an infinite set of commuting conserved
quantities, so that they belong to a hierarchy of integrable systems.

The Benjamin-Ono$_{2}$ hierarchy is a generalization
of these integrable systems \cite{Lebedev1983,Degasperis1991,Degasperis1992},
describing non-linear perturbations and solitonic waves on the edge
of the quantum Hall fluid \cite{abanov2005quantum,Bettelheim2006,Abanov2009,Wiegmann:2011aa},
as well as in further applications of one-dimensional condensed matter
systems \cite{Bettelheim2006,Abanov2009,2012RvMP...84.1253I}. It
is defined in terms of two dynamical fields, $\mathcal{L}(t,\phi)$ and
$\mathcal{J}(t,\phi)$, which we assume to be $2\pi$-periodic in the
$\phi$ coordinate. 

The BO$_{2}$ hierarchy also possesses solitonic solutions and an
infinite set of commuting integrals of motion $H_{j}=\frac{c}{12\pi\ell}\int\mathcal{H}_{j}(\phi)d\phi$,
with $j\in\mathbb{Z}_{>0}$, so that the field equations of the $z$-th
representative can be written in Hamiltonian form as 
\begin{equation}
\dot{\mathcal{L}}=\poisson{\mathcal{L}}{H_{z}},\quad\dot{\mathcal{J}}=\poisson{\mathcal{J}}{H_{z}},\label{eq:hamilton-eqs-2}
\end{equation}
where the Poisson brackets are given by \footnote{In our conventions, the fields are expanded in modes according to
\eqref{eq:modes}, so that $c$ stands for the Virasoro central charge.} 
\begin{equation}
\label{eq:poisson_brackets}
  \begin{aligned}\{\mathcal{L}(\phi),\mathcal{L}(\bar \phi)\} & =-\frac{48\pi}{c}\mathcal{D}_{\phi}\delta\left(\phi-\bar \phi\right),\quad\{\mathcal{L}(\phi),\mathcal{J}(\bar\phi)\}=0,\\[5pt]
\{\mathcal{J}(\phi),\mathcal{J}(\bar \phi)\} & =\frac{6\pi}{c}\partial_{\phi}\delta\left(\phi-\bar\phi\right),
\end{aligned}
\end{equation}
and $\mathcal{D}_{\phi}=\partial_{\phi}{\cal L}+2{\cal L}\partial_{\phi}-2\partial_{\phi}^{3}$.
The first three densities $\mathcal{H}_{j}$ read 
\begin{equation}
\begin{aligned}\mathcal{H}_{1} & =\frac{1}{4}{\cal L}-{\cal J}^{2}\quad,\quad\mathcal{H}_{2}=\frac{2}{3}\left[\frac{1}{4}{\cal L}{\cal J}-\frac{1}{3}{\cal J}^{3}-{\cal J}\mathsf{H}\partial_{\phi}{\cal J}\right],\\[10pt]
\mathcal{H}_{3} & =\frac{1}{8}{\cal L}^{2}-3{\cal L}{\cal J}^{2}-3{\cal L}\mathsf{H}\partial_{\phi}{\cal J}+10\left(\partial_{\phi}{\cal J}\right)^{2}+12{\cal J}^{2}\mathsf{H}\partial_{\phi}{\cal J}+2{\cal J}^{4},
\end{aligned}
\label{eq:classic_Ham}
\end{equation}
where $\mathsf{H}$ is the Hilbert transform defined by the principal
value integral 
\begin{equation}
\mathsf{H}\,F(\phi)=\frac{1}{2\pi}\,\mathcal{P}\int_{0}^{2\pi}F(\bar \phi)\cot\frac{1}{2}(\bar \phi-\phi)\,d\bar \phi.\label{eq:hilbert-transform}
\end{equation}
The remaining conserved quantities of the hierarchy can be obtained
recursively by imposing commutativity,
$\left\{ H_{k},H_{l}\right\} =0$.  More powerful methods to obtain the
integrals of motion can be found in
\cite{Lebedev1983,Degasperis1992,Degasperis1991,Bazhanov1996,Litvinov2013}.
An action principle for the BO$_{2}$ hierarchy can be formally written
in terms of a non-local symplectic form given by the inverse of the
Poisson brackets in \eqref{eq:poisson_brackets}, see
e.g. \cite{gel1979hamiltonian,das1989integrable,olver2000applications}.

For our purposes, it is worth stressing that the BO$_{2}$ equations
are invariant under anisotropic scaling of Lifshitz type with dynamical
exponent $z$ in \eqref{eq:Lifshitz-scaling}, provided that the fields
scale as ${\cal L}\rightarrow\lambda^{-2}{\cal L}$, ${\cal J}\rightarrow\lambda^{-1}{\cal J}$.
Indeed, the conserved charges scale according to $H_{j}\rightarrow\lambda^{-j}H_{j}$
, and thus, the Hamiltonian of the $z$-th representative of the hierarchy,
$H_{z}$, becomes labeled in terms of its scaling dimension $z$,
which matches the dynamical exponent used before. 

In spite of the non--locality introduced through the Hilbert transform,
it is remarkable that BO$_{2}$ can be  quantized.

\subsection{Quantum BO$_{2}$ Hierarchy}

The quantum BO$_{2}$ hierarchy surprisingly emerges in the context
of the AGT correspondence, which describes a relationship between
4d $\mathcal{N}=2$ supersymmetric gauge theories and 2d conformal
field theories~\cite{Alday2010b}. The partition function of certain
type of supersymmetric models, called class-$\mathcal{S}$ models,
is given by the Nekrasov partition function $Z_{\text{inst}}$~\cite{Nekrasov2003a}.
The AGT correspondence states that $Z_{\text{inst}}\propto\mathcal{F}_{c}$,
where $\mathcal{F}_{c}$ is a Liouville CFT conformal block. For the
detailed map between the two sides, see \cite{Alday2010b}. Here,
we just sketch the minimal information about the correspondence that
is useful for our requirements.

The proof of the AGT expansion relies on the introduction of a new
basis of descendant CFT$_{2}$ states, which we call the AFLT basis
\cite{Alba2011}. It starts by considering the tensor product algebra
$\mathcal{A}=\text{Vir}\otimes\mathcal{H}$, spanned by the modes
$L_{n}$ of the Virasoro algebra (Vir) and the modes $a_{n}$ of the
Heisenberg algebra $(\mathcal{H})$, so that $[a_{n},L_{m}]=0$. In
CFT, the generators of $\mathcal{A}$ are given in terms of the energy-momentum
tensor $T$ and the $U(1)$ current $J$. Here we use an alternative
normalization, with respect to CFT, for the mode expansion of these
currents 
\begin{align}
{\hat{{\cal L}}}(\phi) & =\frac{24}{c}\sum_{n=-\infty}^{\infty}L_{n}\,e^{-in\phi}-1,\label{eq:modes}\\[5pt]
{\hat{{\cal J}}}(\phi) & =i\sqrt{\frac{24}{c}}\sum_{n=-\infty,n\neq0}^{\infty}a_{n}\,e^{-in\phi},
\end{align}
to match the conventions set in the classical formulation in section
\ref{sec:classical-bo_2}. As in \cite{Alba2011,Litvinov2013}, we
also discard the zero mode of the $u(1)$ current. Notice that, for
$\hat{\mathcal{L}}$ and $\hat{\mathcal{J}}$ to be Hermitian, we
set $L_{n}^{\dagger}=L_{-n}$ and $a_{n}^{\dagger}=-a_{-n}$.

To proceed, we introduce the standard Liouville notation for the central
charge and conformal dimensions \cite{Ribault:2014aa} 
\begin{equation}
c=1+6Q^{2},\quad Q=b+\frac{1}{b},\quad\Delta(P)=\frac{Q^{2}}{4}-P^{2},\label{eq:liouville-notation}
\end{equation}
where $b$ is the Liouville parameter and the momentum $P$ labels
the primary states.  The orthogonal AFLT basis reads 
\begin{equation}
|P\rangle_{\vec{\lambda}}=\sum_{|\vec{\mu}|=|\vec{\lambda}|}C_{\vec{\lambda}}^{\mu_{1},\mu_{2}}(P)\,a_{-\mu_{1}}L_{-\mu_{2}}|\Delta(P)\rangle,\label{eq:AFLT-basis}
\end{equation}
where the first few coefficients $C_{\vec{\lambda}}^{\mu_{1},\mu_{2}}(P)$
are given in \cite{Alba2011}. Here $\vec{\lambda}=(\lambda_{1},\lambda_{2})$
corresponds to two integer partitions $\lambda_{k}=\{(\lambda_{k})_{1},(\lambda_{k})_{2},\ldots,(\lambda_{k})_{n})\}$,
with $k=1,2$, and $\lambda_{1}\geq\lambda_{2}\geq\cdots\geq\lambda_{n}$.
One of the main conclusions of \cite{Alba2011} is that the insertion
of the completeness relation of the basis \eqref{eq:AFLT-basis} in
a CFT correlator gives the AGT conformal block expansion.

The AFLT basis $\afltbasis$ also diagonalizes an infinite set of
mutually commuting operators $\mathbf{H}_{j}$, $j\in\mathbb{Z}_{>0}$,
given by the quantum BO$_{2}$ integrals of motion and their eigenvalues
can be explicitly obtained \cite{Alba2011,Litvinov2013}. The quantum
integrals of motion $\mathbf{H}_{j}$ lie in the universal enveloping
algebra of $\mathcal{A}$. The first two of them read

\begin{equation}
\begin{aligned} & \mathbf{H}_{1}=\frac{1}{\ell}\left(L_{0}+2\sum_{k=1}^{\infty}a_{-k}a_{k}-\frac{c+1}{24}\right),\\[10pt]
 & \mathbf{H}_{2}=\frac{4i}{3\ell}\sqrt{\frac{6}{c}}\left(\sum_{k=-\infty,k\neq0}^{\infty}a_{-k}L_{k}+2iQ\sum_{k=1}^{\infty}ka_{-k}a_{k}+\frac{1}{3}\sum_{i+j+k=0}a_{i}a_{j}a_{k}\right),
\end{aligned}
\label{eq:bo2-operators}
\end{equation}
while for odd values of $z=2n-1$ one obtains 
\[
\mathbf{H}_{2n-1}=\frac{1}{n\ell}\left(\frac{24}{c}\right)^{n-1}\left(L_{0}^{n}+\cdots\right),
\]
where the ellipsis stands for non-zero modes. Note that we have defined
the operators $\mathbf{H}_{j}$ to be Hermitian, so that the spectrum
is real, and the classical integrals of motion $H_{j}$ obtained from
the densities \eqref{eq:classic_Ham}, can be recovered from \eqref{eq:bo2-operators}
in the semiclassical limit $b\rightarrow0$.

The BO$_{2}$ eigenstates $|P\rangle_{\vec{\lambda}}$ obey $\mathbf{H}_{z}\,|P\rangle_{\vec{\lambda}}=E_{\vec{\lambda}}^{(z)}(P)\,|P\rangle_{\vec{\lambda}},$
with eigenvalues given by $E_{\lambda_{1},\lambda_{2}}^{(z)}(P)=E_{\lambda_{1}}^{(z)}(P)+E_{\lambda_{2}}^{(z)}(-P)$.
For a generic Hamiltonian $\mathbf{H}_{z}$, it was conjectured in
\cite{Alba2011} and \cite{Litvinov2013} that the spectrum can be
written as a sum of two eigenvalues of the Calogero-Sutherland model
plus some extra terms depending only on $\Delta$ and $c$. The Calogero-Sutherland
eigenvalues are given by $h_{\lambda,\mu}^{(z)}=h_{\lambda}^{(z)}(P)+h_{\mu}^{(z)}(-P)$,
where 
\begin{multline}
h_{\lambda}^{(z)}(P)=\frac{2^{z}}{\left(1+z\right)\ell}\left(\frac{6}{b^{2}c}\right)^{\frac{z-1}{2}}\sum_{j=1}^{N_{r}}\left[\left(bP-\frac{b^{2}}{2}+\lambda_{j}+jb^{2}\right)^{z}-\left(bP-\frac{b^{2}}{2}+jb^{2}\right)^{z}\,\right].\label{eq:calogero2}
\end{multline}
We call $h_{\lambda,\mu}^{(z)}$ the \emph{descendant part} of the
spectrum. The descendant part can be obtained from the Bethe ansatz
equations conjectured in \cite{Litvinov2013} and proven by \cite{Feigin:2017gcv}.
We denote the contribution to the energy due to primary states as
$E^{\left(z\right)}(P)$, so that the ground state energy is given
by $E_{0}^{(z)}\equiv E^{\left(z\right)}(\pm\frac{Q}{2})$. The eigenvalues
of the first four operators $\mathbf{H}_{z}$, in our normalization,
are given by 
\begin{equation}
\begin{aligned} & E_{\lambda,\mu}^{(1)}(P)=h_{\lambda,\mu}^{(1)}(P)+E^{(1)}(P),\quad E_{\lambda,\mu}^{(2)}(P)=\frac{1}{2}h_{\lambda,\mu}^{(2)}(P),\\[5pt]
 & E_{\lambda,\mu}^{(3)}(P)=-2h_{\lambda,\mu}^{(3)}(P)-\frac{6}{c\ell}\left(1+b^{2}\right)N_{\lambda,\mu}+E^{(3)}(P),\\[5pt]
 & E_{\lambda,\mu}^{(4)}(P)=\frac{1}{2}h_{\lambda,\mu}^{(4)}(P)+\frac{18}{5}\frac{\left(1+b^{2}\right)}{c}h_{\lambda,\mu}^{(2)}(P),
\end{aligned}
\label{eq:energiesBO2}
\end{equation}
where $N_{\lambda,\mu}=|\lambda|+|\mu|$, and 
\begin{equation}
\begin{aligned}E^{(1)}(P) & =\frac{1}{\ell}\left[\Delta(P)-\frac{c+1}{24}\right],\\[5pt]
E^{(3)}(P) & =\frac{12}{c\ell}\left[\Delta(P)^{2}-\frac{c+5}{12}\Delta(P)+\frac{5c^{2}+52c+15}{2880}\right].
\end{aligned}
\label{eq:primary-energies}
\end{equation}
Note that for a generic heavy state, where $\Delta(P)\approx\left(c/24\right)\mathcal{L}(P)$
in the semiclassical limit, for $z$ odd we have 
\begin{equation}
E^{(z)}(P)=\frac{c}{12\ell}\frac{1}{1+z}[\mathcal{L}(P)]^{\frac{1+z}{2}}.\label{eq:1}
\end{equation}
This corresponds to a classical state with energy $E^{(z)}(P)=\langle\Delta(P)|\hat{H}_{z}|\Delta(P)\rangle$.

Knowing the spectrum then allows us to obtain the leading term of
the entropy in the semiclassical limit. In the case of even values
of $z$, the leading entropy can be obtained along the lines of number
theory, while for odd values of $z$ it can be done through anisotropic
modular invariance.

\emph{Entropy for $z=2n$: } In the semiclassical limit $b\rightarrow0$,
as it occurs for $z=2,4$, we assume that the energy levels have only
contributions from the descendant part, so that $E_{\lambda,\mu}^{(z)}(P)\sim h_{\lambda,\mu}^{(z)}(P)$
to leading order in $c$. From \eqref{eq:calogero2}, if $P\ll b^{-1}$,
we have that 
\begin{equation}
E_{\lambda,\mu}^{(z)}=\varepsilon_{z}\sum_{k}\left(\lambda_{k}^{z}+\mu_{k}^{z}\right).\label{eq:semiclassical-descendant-energies-1}
\end{equation}
This corresponds to energies close to the CFT gap. For states in which
$P\sim\sqrt{c}$, the situation is more complicated, but, for large
enough partitions, the energies are still dominated by \eqref{eq:semiclassical-descendant-energies-1}.
The explicit values of the characteristic energies for $z=2,4$ can
be read from \eqref{eq:energiesBO2} to be given by $\varepsilon_{2}=\frac{2}{3\ell}$
and $\varepsilon_{4}=\frac{8}{5\ell}$.

According to \eqref{eq:semiclassical-descendant-energies-1}, the
asymptotic growth of the number of states then goes as in section
\ref{sec:Microstate-counting-from}, but extended to a two-colored
system. Indeed, the $N$-colored entropy for systems with Lifshitz
scaling in (1+1)--d can be readily obtain from \eqref{eq:Sepsilon}
by the replacement $\varepsilon_{z}\rightarrow\varepsilon_{z}/N^{z}$,
\cite{WIP}. Therefore, in this case the entropy is given by \eqref{eq:Sepsilon}
with $\varepsilon_{z}\rightarrow2^{-z}\varepsilon_{z}$.

\emph{Entropy for $z=2n-1$}: In this case, the energies are no longer
dominated by the descendant part, but instead by the primary part
$E^{(z)}(P)$. In the semiclassical limit, assuming that $\Delta \ll c$, the leading contribution
of $E^{(z)}(P)$ 
comes from normal ordering
of the leading term of the Hamiltonian, $\bm{\mathcal{H}}_{z}\sim{\hat{{\cal L}}}^{\frac{1+z}{2}}$,
so that the ground state energy reads 
\begin{equation}
E_{0}^{\left(z\right)}=\frac{\left(-1\right)^{\frac{1+z}{2}}}{1+z}\frac{c}{12\ell}.\label{eq:groundsBO2}
\end{equation}
Hence, in this case the entropy is determined by \eqref{eq:leading-entropy}
with $E_{0}\left[z\right]=E_{0}^{\left(z\right)}$.

In the next section, we connect the present discussion of the semiclassical
BO$_{2}$ spectrum with gravitation on AdS$_{3}$ and black holes.

\section{Geometrization of Benjamin-Ono$_{2}$ and Black Hole Entropy in 3D}

\label{sec:geom-benj-ono_2}

Following the lines of \cite{Perez2016}, here we show that the BO$_{2}$
hierarchy of integrable systems can be fully geometrized, in the sense
that its dynamics can be equivalently understood in terms of the evolution
of spacelike surfaces and $U(1)$ fields with vanishing field strength
embedded in locally AdS$_{3}$ spacetimes. Let us then consider the
Einstein-Hilbert action with negative cosmological constant in 3D,
endowed with a couple of noninteracting $U(1)$ fields 
\begin{equation}
I=\frac{1}{16\pi G}\int d^{3}x\left[\sqrt{-g}(R+2\ell^{-2})-2\ell\epsilon^{\mu\nu\lambda}\left(A_{\mu}^{+}\partial_{\nu}A_{\lambda}^{+}-A_{\mu}^{-}\partial_{\nu}A_{\lambda}^{-}\right)\right],\label{eq:EH+U1s}
\end{equation}
which agrees with the bosonic sector of $\mathcal{N}=(2,2)$ supergravity
\cite{Achucarro:1989gm}. Note that since the $U(1)$ fields are described
by Chern-Simons actions, the spacetime metric does not acquire a back
reaction due to their presence. Therefore, the field equations imply
that spacetime is of negative constant curvature, carrying two independent
$U(1)$ fields of vanishing field strength.

As done in \cite{Perez2016} (see also \cite{Fuentealba:2017omf}),
it can be shown that there exists a precise set of boundary conditions,
being such that, in the reduced phase space, the field equations obtained
from \eqref{eq:EH+U1s} exactly reduce to (left and right copies of)
BO$_{2}$. This can be seen as follows. According to \cite{Achucarro:1989gm,Witten:1988hc},
up to boundary terms, the action \eqref{eq:EH+U1s} can be written
as the difference of two Chern-Simons actions, both with level $k=\ell/4G$$,$
for independent $SL(2,R)\times U(1)$ gauge fields, so that the dreibein
and the (dualized) spin connection are related to the $SL(2,R)$ gauge
fields as $A^{\pm}$$_{SL(2,R)}=\omega\pm e\ell^{-1}$. We then have
to specify the asymptotic structure of the fields. For simplicity
we restrict the analysis to the left copy, since the extension to
the remaining one is straightforward. It is useful to make a gauge
choice as in \cite{Coussaert:1995zp,Henneaux:2013dra,Bunster:2014mua},
so that the $SL(2,R)\times U(1)$ connection reads 
\begin{equation}
{\cal A}=g^{-1}\left(d+a\right)g\,,\label{eq:Agrande}
\end{equation}
with $g=e^{\log\left(r/\ell\right)L_{0}}$. This gauge choice certainly
simplifies our task, since the remaining analysis can be performed
in terms of the auxiliary gauge field 
\begin{equation}
a=a_{t}dt+a_{\phi}d\phi\;,\label{eq:achico}
\end{equation}
which exclusively depends on $t$, $\phi$.

Thus, the asymptotic form of \eqref{eq:achico} is proposed to be
given by 
\begin{align}
a_{\phi} & =L_{1}-\frac{1}{4}{\cal L}L_{-1}+{\cal J}J_{0},\label{eq:a-phi}\\
a_{t}= & \mu L_{1}-\frac{1}{4}\mu{\cal L}L_{-1}-\mu'L_{0}+\frac{1}{2}\mu''L_{-1}-\frac{1}{8}\xi J_{0},\nonumber 
\end{align}
where $\mu,\xi$ stand for Lagrange multipliers associated to the
dynamical fields ${\cal L},{\cal J}$ respectively. The boundary conditions
then become fully specified only once the Lagrange multipliers are
kept fixed at the boundary, located at a fixed value of the radial
coordinate. Our choice of boundary conditions then consists in precisely
fixing $\mu$ and $\xi$ in terms of the dynamical fields and their
derivatives along $\phi$ according to 
\[
\mu=\frac{48\pi}{c}\frac{\delta H_{z}}{\delta{\cal L}}\quad,\quad\xi=\frac{48\pi}{c}\frac{\delta H_{z}}{\delta{\cal J}},
\]
where $H_{z}$ stands for the $z$-th conserved charge of BO$_{2}$,
with $c$ given by the Brown-Henneaux central charge $c=3\ell/2G$
\cite{Brown1986}.

Since we are dealing with a Chern-Simons theory, the field equations
imply that the $SL(2,R)\times U(1)$ connection ${\cal A}$ is locally
flat, and by virtue of the gauge choice in \eqref{eq:Agrande}, the
field strength of the auxiliary gauge field \eqref{eq:achico} also
vanishes. Therefore, the components of $a$ in \eqref{eq:a-phi} reduce
to an $SL(2,R)\times U(1)$-valued Lax pair formulation of the BO$_{2}$
hierarchy, so that the field equations in \eqref{eq:hamilton-eqs-2}
can be compactly written as

\begin{equation}
f=da+a^{2}=0\,.\label{eq:fchico}
\end{equation}
Therefore, two independent copies of the BO$_{2}$ equations are precisely
recovered from the reduced phase space of the three-dimensional field
equations of \eqref{eq:EH+U1s} endowed with our choice of boundary
conditions.

Furthermore, according to \cite{Perez2016}, the symmetries of the
BO$_{2}$ equations, spanned by the conserved quantities $H_{j}$,
now emerge from the set of diffeomorphisms that preserve the asymptotic
form of the gauge field. Noteworthy, in the geometric framework, the
symmetries of BO$_{2}$ become Noetherian, and hence, the infinite
set of commuting conserved charges $H_{j}$ is precisely obtained
from the corresponding surface integrals in the canonical approach
\cite{Regge:1974zd}\footnote{In the special case of $z=1$ our boundary conditions reduce to the
bosonic part of the ones in \cite{Henneaux:1999ib}, and the asymptotic
symmetry algebra corresponds to two copies of the direct sum of Virasoro
with the Brown-Henneaux central extension and a $u(1)$ current. }. In particular, the total energy of a three-dimensional configuration
that fulfills our boundary conditions, including gravitation and the
$U(1)$ fields, is then given by the sum of left and right Hamiltonians
of BO$_{2}$, i.e., $E=Q[\partial_{t}]=H_{z}^{+}+H_{z}^{-}$.

In sum, the whole structure of classical BO$_{2}$, including its
phase space, the infinite number of commuting charges and its field
equations, emerges from the reduced phase space of gravitation on
AdS$_{3}$ coupled to two $U\left(1\right)$ fields with our boundary
conditions. Hence, this construction provides a gravitational dual
of two noninteracting left and right BO$_{2}$ movers, describing
locally AdS$_{3}$ spacetimes with anisotropic scaling induced by
the choice of boundary conditions. Consequently, any solution of the
BO$_{2}$ equations can be mapped into a locally AdS$_{3}$ spacetime
with suitable $U\left(1\right)$ fields of vanishing field strength.
In particular, one of the most trivial BO$_{2}$ configurations, given
by ${\cal J}^{\pm}=0$ and ${\cal L}^{\pm}=\ell^{-2}\left(r_{+}\pm r_{-}\right)^{2}$
constants, corresponds to the geometry of a BTZ black hole in vacuum
\cite{Banados:1992wn,Banados:1992gq}. Note that in the geometric
picture this configuration is clearly non-trivial because the event
horizon has Hawking temperature and entropy, and its mass and angular
momentum become well defined in terms of left and right BO$_{2}$
energies 
\begin{equation}
H_{z}^{\pm}\left[{\cal L}_{\pm}\right]=\frac{c}{12\ell}\frac{1}{1+z}{\cal L}_{\pm}^{\frac{1+z}{2}},\label{eq:BTZenergy}
\end{equation}
provided that $z=2n-1$. Note that \eqref{eq:BTZenergy} agrees with
\eqref{eq:1}.

This geometric realization suggests that the reduced
gravitational phase space could be
quantized in terms of two copies of BO$_{2}$, so that the states would
be given by the AFLT ones in \eqref{eq:AFLT-basis}.
Indeed, two points are worth to be emphasized:

(i) The ground state energy of quantum BO$_{2}$ in the semiclassical
limit, given by $E_{0}^{(z)}$ in \eqref{eq:groundsBO2}, exactly
coincides with the one of the geometric configuration of lowest energy,
determined by global AdS$_{3}$ spacetime. Indeed, left and right
energies of AdS$_{3}$ correspond to \eqref{eq:BTZenergy} with ${\cal L}_{\pm}=-1$,
and hence 
\begin{equation}
\underbrace{H_{z}^{\pm}[-1]}_{\text{AdS}_{3}}=\underbrace{\:\:E_{0}^{(z)}\:\:}_{\text{Quantum }\text{\text{BO}}_{2}}=\frac{\left(-1\right)^{\frac{1+z}{2}}}{1+z}\frac{c}{12\ell}.\label{eq:gsenergy}
\end{equation}

(ii) The leading term of the asymptotic growth of the number of states
is then obtained from \eqref{eq:leading-entropy} for both copies,
i.e., 
\begin{equation}
S=2\pi\ell(1+z)\left[\left(\frac{\left|E_{0}^{+}[z^{-1}]\right|}{z}\right)^{\frac{z}{1+z}}E_{+}^{\frac{1}{1+z}}+\left(\frac{\left|E_{0}^{-}[z^{-1}]\right|}{z}\right)^{\frac{z}{1+z}}E_{-}^{\frac{1}{1+z}}\right],\label{eq:Cardy}
\end{equation}
where $E_{0}^{+}[z]=E_{0}^{-}[z]$ stand for left and right energies
of the ground state, determined by \eqref{eq:gsenergy}. Hence, for
left and right energies given by the ones of the black hole, i.e.,
$E_{\pm}=H_{z}^{\pm}\left[{\cal L}_{\pm}\right]$ in \eqref{eq:BTZenergy},
noteworthy, the entropy obtained from the anisotropic extension of
Cardy formula \eqref{eq:Cardy} exactly reduces to the one of Bekenstein
and Hawking, given by 
\[
S=\frac{A}{4G}.
\]

\acknowledgments{ The authors thank valuable discussions with Sebas
  Eliens, Hernán González, Rodrigo Pereira, Miguel Pino, Pablo Rodríguez, David
  Tempo, Jacopo Viti, Paul Wiegmann and Alexander
  B. Zamolodchikov. The work of DM was supported by the grant
  No. 16-12-10344 of the Russian Science Foundation. FN thanks Máté
  Lencsés for pointing out the Hardy-Ramanujan paper. FN also thanks
  Jun'ichi Shiraishi and, specially, Yohei Tutiya for the initial
  discussions on this project and kind hospitality at the Komaba
  Mathematics section of the University of Tokyo, where part of this
  work was developed. FN thanks the organizers of the
  \emph{Latin-American Workshop on Gravity and Holography} held in São
  Paulo in June, 2018 for the opportunity to present the main results
  of this work and for financial support. FN acknowledges the
  Brazilian Ministry of Education for the financial support. AP thanks
  Stefan Theisen for his kind hospitality at the MPI für
  Gravitationsphysik in Golm, and the German Academic Exchange Service
  (DAAD) for financial support through the ``Re-invitation Programme
  for Former DAAD Scholarship Holders''. AP and RT thank Fondecyt grants Nº 1161311, 1171162, 1181031 and
  1181496 for financial support. The Centro de Estudios Científicos (CECs) is funded by the
  Chilean Government through the Centers of Excellence Base Financing
  Program of Conicyt. }

 \bibliographystyle{JHEP}
\bibliography{bcs-1}

\end{document}